\documentclass[12pt]{JHEP3}

\let\ifpdf\relax
\usepackage{ifpdf}
\usepackage{graphics}
\usepackage{graphicx}
\usepackage{amssymb}
\usepackage{amsmath}
\usepackage{slashed}
\usepackage{cite}
\usepackage{epsfig}
\usepackage{datetime}

\newcommand{\be}{\begin{equation}}
\newcommand{\ee}{\end{equation}}
\newcommand{\bal}{\begin{align}}
\newcommand{\eal}{\end{align}}
\newcommand{\bea}{\begin{eqnarray}}
\newcommand{\eea}{\end{eqnarray}}

\def\Tr{{\rm Tr}}

\title{Large distance expansion of Mutual Information for disjoint disks in a free scalar theory}

\author{Cesar A. Ag\'on, Isaac Cohen-Abbo and Howard J. Schnitzer  \\
{\small  Martin Fisher School of Physics, Brandeis University, \\ \ \ \ \ \ Waltham, MA 02454, USA\\
}

E-mail: \email{caagon87@brandeis.edu},\\ \email{icohen@brandeis.edu},
 \email{schnitzr@brandeis.edu}}

\preprint{
 BRX-TH-6295}

\abstract{

We compute the next-to-leading order term in the long-distance expansion of the mutual information for free scalars in three space-time dimensions. The geometry considered is two disjoint disks separated by a distance $r$ between their centers.  No evidence for non-analyticity in the R\'enyi parameter $n$ for the continuation $n \rightarrow 1$ in the next-to-leading order term is found.
}

\begin{document}

\section{Introduction}
 Cardy \cite{Cardy:2013nua} has presented a general framework for the mutual R\'enyi information of two disjoint compact spatial regions $A$ and $B$ for a $d+1$ dimensional conformal field theory in the limit when the separation $r$ between $A$ and $B$ is much greater than the sizes $R_A $ and $R_B$. The method involves the replica trick \cite{Holzhey:1994we,Calabrese:2004eu} for computing the R\'enyi entropies $S_A^{(n)}=(1-n)^{-1}\log\Tr \rho_A^{n}$, and from these quantities the von Neumann entropy $S_A=-\Tr \rho_A \log \rho_A$ of the reduced density matrix $\rho_A$, obtained as 
\bea
&& S_A=\lim_{n\to 1} S_{A}^{(n)}\,.
\eea
This limit generally involves a non-trivial analytic continuation of the R\'enyi parameter $n$.

 Similarly, from the R\'enyi mutual information 
\bea
I^{(n)}(A,B)\equiv S_A^{(n)}+S_B^{(n)}-S_{A\cup B}^{(n)}
\eea
one can obtain its corresponding mutual information
\bea
I(A,B)\equiv S(A)+S(B)-S(A\cup B)\,,
\eea
taking the same limit.

For a free scalar theory in $(d+1)$ dimensions, one has in general the leading term in the large $r$ expansion
\bea
\label{renyinfo}
I^{(n)}(A,B)\sim g_d^{(n)} \left(\frac{R_A R_B}{r^2} \right)^{d-1}\,.
\eea
For the simple case where $A$ and $B$ are spheres of radii $R_A$ and $R_B$, Cardy found for $3+1$ dimensions
\bea
g_3^{(n)}=\frac{n^4-1}{15n ^3 (n-1)} ,
\eea

\bea
g_2^{(1)}=\frac{1}{3}
\eea
for $2+1$ dimensions \footnote{Recently, the  coefficient $g_{d}^{(1)}$ for any dimension was found to be
$g_{\Delta}^{(1)}={\cal N}_\Delta \frac{\sqrt{\pi} \Gamma(2\Delta +1)}{4\Gamma (2\Delta+\frac 32)}\,,
$
for arbitrary CFTs \cite{Agon:2015ftl}. Here $\Delta$ is the lowest scaling dimension of the CFT operators and  ${\cal N}_\Delta$ its degeneracy. Notice that for free CFTs, the $d$ dependence comes from the relation between $\Delta$ and $d$} and a known definite integral for $g_2^{(n)}$. 

A compound system with a cusp, consisting of two identical spheres in contact widely separated from two other identical spheres in contact also gives a mutual R\'enyi entropy  (\ref{renyinfo}) with\cite{Schnitzer:2014zva}
\bea
g_d^{(n)}&=& 2n\left[\sum_{l=1}^{\infty} \frac{(-1)^{l+1}}{l^{d-1}}\right]^2\nonumber \\
&&\left\{ \frac14\sum_{j=1}^{n-1}(\frac1j)^2 +\frac{(n-1)^2}{n^2}\right\}(n-1)^{-1} \,,
\eea
where now in  (\ref{renyinfo}) $R_A$ is the radius of one of the spheres in $A$ and similarly for $B$. Thus, the leading term of the expansion  is of universal form, with $g_d^{(n)}$ encoding the geometry of $A$ and $B$.

For $d \geq 2$, higher order terms in the large distance expansion of the mutual information $I(A,B)$ requires the contribution of the stress tensor $T_{\mu \nu}$ and current $\partial_{\mu}\phi$ to evaluate the corresponding coefficients. For example, for $d=2$, which is the subject of this work, contributions from $T_{\mu \nu}$ and $\partial_{\mu}\phi$ only appear starting at the next-to-next to leading order $\sim \left(\frac{R_A R_B}{r^2}\right)^3$ and their contribution is of equal importance as that of $\Phi_j$ (analogous to the one computed in Section 4).
For $d=3$, this additional terms contribute already in the next to leading order term and their contribution dominate for $d>3$.

Care must be taken in these calculations to ensure that spurious singularities do not appear in the many integrations to be carried out. Given the complexity of the calculation, one may be concerned with a possible non-analyticity in the R\'enyi parameter $n$. However, no evidence for such non-analyticity is found in the continuation $n \rightarrow 1$ of the next to leading order term in $1/r$ for free scalar fields in $3$ space-time dimensions ($d=2$).

For $d=2$  free scalar fields, we have the long distance expansion
\bea
\label{subs}
 \left(\frac{R_A R_B}{r^2}\right);\left(\frac{R_A R_B}{r^2}\right)^2; \left(\frac{R_A R_B}{r^2}\right)^3; \left(\frac{R_A R_B}{r^2}\right)^4; \cdots ;
\eea
where the sub-leading term will be the subject of our analysis.
It is found  that
\bea
\label{result}
I(A,B)\sim \frac13\left(\frac{R_AR_B}{r^2}\right)+\frac13\left(\frac{R_AR_B}{r^2}\right)^2\left(\frac{6}{5}+\frac{4}{\pi^2}\right)\,.
\eea
The appearance of $(1/\pi^2)$ in (\ref{result}) is unexpected.

\section{Review of Cardy's paper \cite{Cardy:2013nua}}
Consider two disjoint compact objects $A$ and $B$ in a CFT, It was shown in \cite{Headrick:2010zt} \cite{Calabrese:2009ez} \cite{Calabrese:2010he}  for 1+1 dimensions, that the R\'enyi mutual information given by
\bea
I^{(n)}(A,B)\equiv S_A^{(n)}+S_B^{(n)}-S_{A\cup B}^{(n)} \, ,
\eea
has the following expansion 
\bea
\label{expansion1/r}
I^{(n)}(A,B)=\sum_{\{k_j\}}C_{A}^{(n)}(\{ k_j\})C_{B}^{(n)}(\{k_j\})\left(\frac{R_{A}R_{B}}{r^2}\right)^{\sum_{l=1}^n x_{k_l}}\,,
\eea
where the $x_{k_l}$ are the scaling dimension of the operators $\Phi_{k_l}$. This expansion was studied in detail in \cite{Calabrese:2010he}. Cardy generalized the equation (\ref{expansion1/r}) for higher space-time dimensions, and it is the intent of this section to present a brief review of his arguments.

The R\' enyi entropy for a given region $X$ is given by
\bea
S^{(n)}_X=\frac{1}{1-n}\log \Tr_{{\cal H}_X}\rho^n_X \, ,
\eea
which in turn can be computed in terms of a path-integral on a conifold
\bea
S^{(n)}_X=\frac{1}{1-n}\log\left(\frac{Z({\cal C}^{(n)}_X)}{Z^n}\right)\,.
\eea

Here $Z$ is the partition function on the original space, and $Z({\cal C}^{(n)}_X)$ is the $n^{th}$ associated conifold. The R\'enyi entropies are then
\bea
\label{renyiinfo}
I^{(n)}(A,B)=\frac{1}{1-n}\log\left(\frac{Z({\cal C}^{(n)}_{A\cup B})Z^{n}}{Z({\cal C}^{(n)}_A)Z({\cal C}^{(n)}_B)}\right)\,.
\eea
The basic idea that leads to the expansion (\ref{expansion1/r}) is the observation that the sewing operation on the entangling regions can be thought of by a distant observer as a semi-local operation, and so  can be implemented through a weighted sum of a product of local operators. That is
\bea
\frac{Z({\cal C}^{(n)}_{A\cup B})}{Z^{n}}=\langle  \Sigma_A^{(n)} \Sigma_B^{(n)} \rangle\,,
\eea
where
\bea
\Sigma_{A}^{(n)}=\frac{Z({\cal C}_A^{(n)})}{Z^n}\sum_{\{k_j\}}C_{\{k_j\}}^{A} \prod_{j=0}^{n-1} \Phi_{k_j}(r^{j}_A)\,.
\eea
The coefficients of the expansion can be obtained by
\bea
C_{\{k_j\}}^{A}=\lim_{\{r^{(j)}\}\to \infty}|r^{(j)}|^{\sum_j 2x_{k_j}}\langle \prod_{j} \Phi_{k_j}(r^{j})\rangle_{{\cal C}^{(n)}_A}
\eea
and the ratio of partition functions relevant for the evaluation of the mutual information is
\bea
\label{insidelog}
\frac{Z({\cal C}^{(n)}_{A\cup B})Z^n}{Z({\cal C}^{(n)}_{A})Z({\cal C}^{(n)}_{B})}=\sum_{\{k_j\}}{ C}^A_{\{k_j\}}{C}^B_{\{k_j\}}r^{-2\sum_j x_{k_j}}\,.
\eea

One can write the left-hand side of (\ref{insidelog}) as
\bea
\label{Pn}
P^{(n)}\equiv \frac{P_{A\cup B}^{(n)}}{P_A^{(n)}P_B^{(n)}}\, ,
\eea
where $P_{X}^{(n)}=Z({\cal C}^{(n)}_{X})/Z^n$,
and expand (\ref{renyiinfo}) systematically in the large $r$ expansion;
 so that
\bea
\label{expansioninfo}
I^{(n)}(A,B)=\frac{1}{n-1}\{  [P^{(n)}]_{(1)} - \frac 12[P^{(n)}]_{(1)}^2+ [P^{(n)}]_{(2)}+\cdots \}\,,
\eea
where $[P^{(n)}]_{(m)}$ corresponds to the order $m$ term in the expansion of (\ref{Pn}), that is $[P^{(n)}]_{(m)}\sim (R_AR_B/r^2)^m$. The  $[P^{(n)}]_{(0)}$ term is equal to one, as it is obtained from the contribution of identity operator $\Phi_{k_j}=\mathbf{1}$, and does not contribute to (\ref{expansioninfo}).

 If one is only interested in the mutual information, that is
\bea
I(A,B)= \lim_{n\to 1} I^{(n)}(A,B)
\eea
 one can ignore the second term in (\ref{expansioninfo}) as $[P^{(n)}]_{(1)}$ is  ${\cal O}(n-1)$ and therefore $[P^{(n)}]_{(1)}^2$ is ${\cal O}(n-1)^2$. Also each term $[P^{(n)}]_{(m)}$ for $m\geqq 2$ should go to zero at least as ${\cal O}(n-1)^2$ when $n$ goes to $1$.

\section{Leading contribution to mutual information}

Cardy \cite{Cardy:2013nua} computed the leading long distance contribution to the mutual information of widely separated disks. In this section we will briefly review his calculation. The lowest order coefficients ${C}^A_{\{k_j\}}$ are
\bea
C_{jj'}^A\equiv C_{0,...,1,...,1,...0}=\lim_{x_1,x_2\to \infty}(x_1 x_2)\langle \phi_j(x_1)\phi_{j'}(x_2)\rangle_{{\cal C}^{(n)}_A}
\eea
for $j\neq j'$ and
\bea
C_{jj}^A\equiv C_{0,...,2,...0}=2^{-1/2}\lim_{x\to \infty}x^{2}\langle :\phi^2_j(x):\rangle_{{\cal C}^{(n)}_A}\,.
\eea
In the case of spheres we can use a conformal transformation that takes the spherical surface of interest (say one with radius $R_A$) in to an infinite plane $\mathbb{R}^{d-1}$, and therefore the original conifold is transformed to ${\cal C'}_A^{(n)}=\{$2-dimensional cone of opening angle $2\pi n\}\times \mathbb{R}^{d-1}$. This transformation also takes points at infinity and maps them to points on a unit sphere so that the previous coefficients in terms of this new coordinates becomes
\bea
C_{jj'}^A&=&2R_A\langle \phi_j(1)\phi_{j'}(1)\rangle_{{\cal C'}^{(n)}_A}\,, \nonumber \\
C_{jj}^A&=&2^{-1/2}R_A\langle :\phi_j^2(1):\rangle_{{\cal C'}^{(n)}_A}\,.
\eea
The Green functions for free scalar fields for $d=2$, were constructed using the method of images and an analytic continuation in $n$. The final result is
\bea
G^{(n)}(1,\theta,0)=\frac{1}{2\pi}\int_0^{\infty}\frac{x^{(\theta/2\pi)-1}(1-x)}{(1+x)(1-x^{n})}dx\,.
\eea
In terms of the new coordinates
\bea
\langle \phi_j(1)\phi_{j'}(1)\rangle_{{\cal C}^{(n)}_A}=\frac{1}{2\pi}\int_0^{\infty}\frac{x^{|j-j'|-1}(1-x)}{(1+x)(1-x^{n})}dx\,.
\eea

These coefficients are what is needed  to evaluate the ratio of the partition function (\ref{insidelog}), so the leading term is
\bea
r^{-2}(\frac 12 \sum_{j\neq j'}C_{jj'}^A C_{jj'}^B+\sum_{j}C_{jj}^A C_{jj}^B)\, ,
\eea
which for the R\'enyi mutual information implies \footnote{The symbol $\sim$ accounts for the fact that the RHS of (\ref{sim}) is the leading large distance term for $n$ close to $1$.} \cite{Cardy:2013nua}
\bea\label{sim}
I^{(n)}(A,B)&\sim&\frac{1}{n-1}\frac{2R_A R_B}{r^2}\Big(\sum_{j\neq j'}\langle \phi_j(1)\phi_{j'}(1)\rangle_{{\cal C}^{(n)}_A}\langle \phi_j(1)\phi_{j'}(1)\rangle_{{\cal C}^{(n)}_B}\nonumber \\
&&\qquad +\sum_{j}\langle :\phi_j^2(1):\rangle_{{\cal C}^{(n)}_A}\langle :\phi_j^2(1):\rangle_{{\cal C}^{(n)}_A}\Big)\,.
\eea
We can fix one index, and put an overall factor of $n$ using the cyclic symmetry of the sum
 \bea
 \label{sym}
I^{(n)}(A,B)&\sim&\frac{n}{n-1}\frac{2R_A R_B}{r^2}\Big(\sum_{j=1}^{n-1}\langle \phi_j(1)\phi_{0}(1)\rangle_{{\cal C}^{(n)}_A}\langle \phi_j(1)\phi_{0}(1)\rangle_{{\cal C}^{(n)}_B}\nonumber \\
&&\qquad +2\langle :\phi_0^2(1):\rangle_{{\cal C}^{(n)}_A}\langle :\phi_0^2(1):\rangle_{{\cal C}^{(n)}_A}\Big)\,.
\eea
Since we are interested in the mutual information, we focus on the terms that survive the $n\to 1$ limit. Taking into account the fact that $\langle :\phi_0^2(1):\rangle \sim O(n-1)$, it is clear that those terms do not contribute to (\ref{sym}) and can be safely dropped. The remaining sum is  
\bea
\label{leadingsingle}
\sum_{j=1}^{n-1}\langle \phi_j(1)\phi_{0}(1)\rangle_{{\cal C}^{(n)}_A}&& \!\!\! \langle \phi_j(1)\phi_{0}(1)\rangle_{{\cal C}^{(n)}_B}\nonumber \\
&&=\frac{1}{4\pi^2}\int_0^{\infty}\int_0^{\infty}\frac{(1-(xy)^{n-1})(1-x)(1-y)}{(1+x)(1+y)(1-xy)(1-x^n)(1-y^n)}dxdy\nonumber \\
&& \sim -\frac{n-1}{4\pi^2}\int_0^{\infty}\int_0^{\infty}\frac{\log(xy)}{(1+x)(1+y)(1-xy)}dxdy \,.
\eea

The second line in (\ref{leadingsingle}) is obtained from the first  when the $n\to 1$ limit is taken. This integral can be done by a change of variables $y'=y$ and $x'=xy$ which transforms it to
\bea
-\frac{n-1}{4\pi^2}\int_0^{\infty}dx'\frac{\log(x')}{(1-x')}\int_0^{\infty}\frac{dy'}{(1+y')(x'+y')}&=&\frac{n-1}{4\pi^2}\int_0^{\infty}dx'\frac{(\log(x'))^2}{(1-x')^2}\nonumber \\
&=&\frac{(n-1)}{6} \, .
\eea
The mutual information can be evaluated  by expanding the $\log$ in (\ref{renyiinfo}) and taking $n\to 1$ limit,  that is

 \bea
I(A,B)&\sim&\lim_{n\to 1}\frac{n}{n-1}\frac{2R_A R_B}{r^2}\frac{(n-1)}6=\frac13\left(\frac{R_A R_B}{r^2}\right)\,.
\eea

\section{Next to leading term}

In this section we apply the procedure of the previous section to evaluate the next to leading order term in the expansion (\ref{expansioninfo}). 

In terms of the $C^X_{j\cdots j'}$ coefficients, the next to leading term  is
\bea
\label{subleading}
r^{-4}(\frac 12 \sum_{j\neq j'}C^A_{jjj'j'}C^B_{jjj'j'}+\frac{1}{3!}\sum_{j\neq j'\neq l}C^A_{jj'll}C^B_{jj'll}+\frac1{4!}\sum_{j\neq j'\neq l\neq l'}C^A_{jj'll'}C^B_{jj'll'})\,.
\eea
Using the same argument as above  we can do one of the sums explicitly due to the cyclic symmetry of the labels
\bea
\label{subleadingn}
\frac{n}{r^{4}}(\frac 12 \sum_{j=1}^{n-1}C^A_{jj00}C^B_{jj00}+\frac{1}{3!}\sum_{j\neq j'}^{n-1}C^A_{jj'00}C^B_{jj'00}+\frac1{4!}\sum_{j\neq j'\neq l}^{n-1}C^A_{jj'l0}C^B_{jj'l0})\,.
\eea
We would like to evaluate each of these terms individually, considering only the ${\cal O}(n-1)$ contributions, which will be the only ones which eventually will contribute to the mutual information.

The coefficients appearing in (\ref{subleadingn}) in terms of correlators of  the fundamental fields $\phi_j$, and $\Phi_j$, are given by \footnote{For notational convenience, here after we use simple brackets $\langle \rangle$ instead of the more accurate $\langle \rangle_{{\cal C}_A}$.}
\bea
C_{jj00}^A&=&(2R_A)^2\langle \Phi_j(1)\Phi_0(1)\rangle \, , \nonumber \\ 
C_{jj'00}^A&=&(2R_A)^2\langle \phi_j(1)\phi_{j'}(1)\Phi_0(1)\rangle \,, \nonumber  \\ 
C_{jj'l0}^A&=&(2R_A)^2\langle \phi_j(1)\phi_{j'}(1)\phi_l(1)\phi_0(1)\rangle\,,
\eea
 where $\Phi_j(1)\equiv \frac 1{\sqrt{2}}:\phi_j^2(1):$. 
 
The expression in (\ref{subleadingn}) contains terms that differ in the number of nested sums involved. The higher this number is the more complex the evaluation turns out to be. In the rest of the paper we will evaluate each of this terms

\subsection{Evaluation of $\frac{n}{2r^4}\sum_{j=1}^{n-1}C^A_{jj00}C^B_{jj00}$}
Using Wick's theorem the coefficient 
\bea
\label{44}
C_{jj00}^A&=& 2R^2_A\langle :\phi_j^2(1)::\phi_{0}^2(1):\rangle \nonumber \\
&=&2R^2_A(\langle :\phi_j^2(1):\rangle \langle:\phi_{0}^2(1):\rangle+2\langle\phi_j(1)\phi_0(1)\rangle \langle\phi_j(1)\phi_0(1)\rangle)\,.
\eea
In the $n\to 1$ limit, any one point function operator in the conifold of singularities goes to zero at least as ${\cal O}(n-1)$. As it will be evident later, any sum gives rise to an $(n-1)$ factor. Therefore, we can safely neglect any term which is already ${\cal O}(n-1)$ or higher before doing a sum, in our case at hand we can ignore the term $\langle:\phi_{j}^2(1):\rangle\langle:\phi_{0}^2(1):\rangle$ in (\ref{44}). 

That means that our term of interest is
\bea
\frac n{2r^4}\sum_{j=1}^{n-1}C^A_{jj00}C^B_{jj00}=\frac{8n R^2_AR^2_B}{r^4}\sum_{j=1}^{n-1}\langle\phi_j(1)\phi_0(1)\rangle^4\,.
\eea
 This term can be evaluated using the same technique used in the evaluation of the leading contribution (\ref{leadingsingle}). This entails the use of the explicit integral representation of the two point, and carrying out the sum explicitly
 \bea\label{4.6}
 &&\sum_{j=1}^{n-1}\langle\phi_j(1)\phi_0(1)\rangle^4 \nonumber \\
 &&=\frac{1}{16\pi^4}\int_0^\infty... \int_0^\infty\frac{(1-(xyzw)^{n-1})(1-x)(1-y)(1-z)(1-w)dxdydzdw }{(1+x)(1+y)(1+z)(1+w)(1-xyzw)(1-x^n)(1-y^n)(1-z^n)(1-w^n)}\,, \nonumber \\
 \eea
 which in the $n\to 1$ limit evaluates to \footnote{See Appendix (\ref{A1}) for the details of the evaluation}
 \bea
 &&\sum_{j=1}^{n-1}\langle\phi_j(1)\phi_0(1)\rangle^4 =\frac{(n-1)}{30}
 \eea
 and so we conclude that 
 \bea
\frac n{2r^4}\sum_{j=1}^{n-1}C^A_{jj00}C^B_{jj00}=\frac{4n(n-1)}{15}\left(\frac{R_AR_B}{r^2}\right)^2\,.
\eea
 
\subsection{Evaluation of $\frac{n}{3!\, r^4}\sum_{j\neq j'}C^A_{jj'00}C^B_{jj'00}$} 
The second term in (\ref{subleadingn}) requires the evaluation of a double sum and therefore some extra care should be taken. 
First, we need $C^{A,B}_{jj'00}$
\bea
C_{jj'00}^A=(2R_A)^2\frac1{\sqrt{2}}\langle \phi_j(1)\phi_{j'}(1):\phi^2_0(1):\rangle\,,
\eea
which can be rewritten as
\bea
C_{jj'00}^A=4R^2_A\frac1{\sqrt{2}}(\langle \phi_j(1)\phi_{j'}(1)\rangle \langle:\phi_{0}^2(1):\rangle+2\langle\phi_j(1)\phi_0(1)\rangle \langle\phi_{j'}(1)\phi_0(1)\rangle)
\eea
using Wick's theorem.

Here again we neglect the term which contains the one point function $\langle:\phi_{0}^2(1):\rangle$ for the same reason given above, and so we have
\bea
\frac n{3!r^4}\sum_{j\neq j'}^{n-1}C^A_{jj'00}C^B_{jj'00}&=&\frac{16n R^2_AR^2_B}{3r^4}\sum_{j\neq j'}^{n-1}\langle\phi_{j'}(1)\phi_0(1)\rangle^2\langle\phi_j(1)\phi_0(1)\rangle^2\nonumber \\
&=&\frac{32n R^2_AR^2_B}{3r^4}\sum_{j=2}^{n-1}\langle\phi_{j}(1)\phi_0(1)\rangle^2\sum_{j'=1}^{j-1}\langle\phi_{j'}(1)\phi_0(1)\rangle^2\,,
\eea
where in the second line we have separated the sum in two, each one with a different order: $j>j'$ and $j'>j$, and used the symmetry $j\leftrightarrow j'$ to equate the two terms and add them up. 
\bea \label{4.12}
&&\sum_{j=2}^{n-1}\langle\phi_{j}(1)\phi_0(1)\rangle^2\sum_{j'=1}^{j-1}\langle\phi_{j'}(1)\phi_0(1)\rangle^2\nonumber \\
&&=\frac{1}{16\pi^4}\sum_{j=2}^{n-1}\int_0^{\infty}\frac{dxdydzdw (1-(zw)^{j-1})(xy)^{j-1}(1-x)(1-y)(1-z)(1-w)}{(1-zw)(1+x)(1+y)(1+z)(1+w)(1-x^n)(1-y^n)(1-z^n)(1-w^n)}\,,\nonumber \\
\eea
in the $n\to 1$ limit, we can evaluate exactly this expression as shown in Appendix (\ref{A2}), where we find that 
\bea
&&\sum_{j=2}^{n-1}\langle\phi_{j}(1)\phi_0(1)\rangle^2\sum_{j'=1}^{j-1}\langle\phi_{j'}(1)\phi_0(1)\rangle^2=-\frac{(n-1)}{60}\,.
\eea
Therefore, the net contribution from the double sum term is
\bea
\frac n{3!r^4}\sum_{j\neq j'}^{n-1}C^A_{jj'00}C^B_{jj'00}=-\frac{8n(n-1)}{45}\left(\frac{R_AR_B}{r^2}\right)^2\,.
\eea
\subsection{Evaluation of $\frac{n}{4! r^4}\sum_{j\neq j'\neq l}^{n-1}C^A_{jj'l0}C^B_{jj'l0}$}
The final term we want to evaluate involves the triple sum
\bea
\label{triplesum}
\frac{n}{ 4!r^4}\sum_{j\neq j'\neq l}^{n-1}C^A_{jj'l0}C^B_{jj'l0}\,.
\eea
As before,we rewrite higher point function into lower ones using Wick's theorem 
\bea
C^A_{jj'l0}&=&4R_A^2\langle \phi_j(1)\phi_{j'}(1)\phi_l(1)\phi_0(1) \rangle \nonumber \\
&=&4R_A^2(\langle \phi_j(1)\phi_{j'}(1)\rangle \langle \phi_l(1)\phi_0(1) \rangle+\nonumber \\
&& \quad \quad \langle \phi_j(1)\phi_{l}(1)\rangle \langle \phi_{j'}(1)\phi_0(1) \rangle+\langle \phi_j(1)\phi_{0}(1)\rangle \langle \phi_{j'}(1)\phi_l(1) \rangle]\,.
\eea
Thus (\ref{triplesum}) is
\bea
\frac{16n}{ 4!}\left(\frac{R_AR_B}{r^2}\right)^2&&\Big[\sum_{j\neq j'\neq l}^{n-1}3\langle \phi_j(1)\phi_{j'}(1)\rangle^2\langle \phi_l(1)\phi_0(1)\rangle^2+\nonumber \\&& +\sum_{j\neq j'\neq l}^{n-1}6 \langle \phi_0(1)\phi_j(1)\rangle \langle \phi_j(1)\phi_l(1) \rangle\langle \phi_{j'}(1)\phi_{l}(1)\rangle \langle \phi_{j'}(1)\phi_0(1) \rangle
\Big]\nonumber \\
\eea
after relabelling   the indices to identify equal terms. The full expression preserves the symmetry $j\leftrightarrow j'$ so we can give an order to these two indices ($j>j'$) and multiply the final expression by two. To evaluate this we have to sum over all possible orders that respect the given order ($j>j'$). That means we will have three different terms from each sum, those are: 1) $j>j'>l$, $\,$  2) $l>j>j'$ and 3) $j>l>j'$. To avoid confusion let use $k$ instead of $j'$. The triple sum terms are
\bea \label{4.18}
&& 4n\left(\frac{R_AR_B}{r^2}\right)^2\sum_{j=3}^{n-1}\sum_{k=2}^{j-1}
\sum_{l=1}^{k-1}\Big[\langle \phi_j(1)\phi_k(1)\rangle^2\langle \phi_l(1)\phi_0(1)\rangle^2+\langle \phi_k(1)\phi_l(1)\rangle^2\langle \phi_j(1)\phi_0(1)\rangle^2\nonumber \\
&& \qquad\qquad \qquad +\langle \phi_j(1)\phi_l(1)\rangle^2\langle \phi_k(1)\phi_0(1)\rangle^2\nonumber \\
&&\qquad \qquad\quad \quad +2\langle \phi_0(1)\phi_j(1)\rangle \langle \phi_j(1)\phi_l(1) \rangle\langle \phi_k(1)\phi_{l}(1)\rangle \langle \phi_k(1)\phi_0(1) \rangle\nonumber \\ 
&& \qquad\qquad \qquad \quad+2\langle \phi_0(1)\phi_k(1)\rangle \langle \phi_k(1)\phi_j(1) \rangle\langle \phi_j(1)\phi_{l}(1)\rangle \langle \phi_l(1)\phi_0(1) \rangle\nonumber \\
&&\qquad \qquad\qquad \qquad +2\langle \phi_0(1)\phi_l(1)\rangle \langle \phi_k(1)\phi_l(1) \rangle\langle \phi_j(1)\phi_{k}(1)\rangle \langle \phi_j(1)\phi_0(1) \rangle
\Big]\,. \nonumber \\
\eea
Writing all propagators in terms of integrals, and carrying out the sums in a convenient order, we find a similar looking expression to the ones we have evaluated in Sec 2, with the final answer
\bea
\frac{n}{ 4!r^4}\sum_{j\neq j'\neq l}^{n-1}C^A_{jj'l0}C^B_{jj'l0}=
{4n(n-1)}\left(\frac{R_AR_B}{r^2}\right)^2\left(\frac{1}{30}+\frac{2}{45}+\frac{1}{3\pi^2}\right)\,.
\eea
For the specifics of this evaluation the reader is advised to look at Appendix (\ref{A3}).

Adding all the evaluated terms, that is the terms with single, double and triple sums in (\ref{subleadingn}), we get
\bea
&&\frac{n}{r^{4}}(\frac 12 \sum_{j=1}^{n-1}C^A_{jj00}C^B_{jj00}+\frac{1}{3!}\sum_{j\neq j'}^{n-1}C^A_{jj'00}C^B_{jj'00}+\frac1{4!}\sum_{j\neq j'\neq l}^{n-1}C^A_{jj'l0}C^B_{jj'l0})\nonumber \\
&&\qquad \qquad \qquad =\frac{n(n-1)}{3}\left(\frac{R_AR_B}{r^2}\right)^2\left(\frac{6}{5}+\frac{4}{\pi^2}\right)\,.
\eea
The mutual information to second order in the long distance expansion parameter $\left(\frac{R_AR_B}{r^2}\right)$ is then given by:
\bea
I(A,B)\sim \frac13\left(\frac{R_AR_B}{r^2}\right)+\frac13\left(\frac{R_AR_B}{r^2}\right)^2\left(\frac{6}{5}+\frac{4}{\pi^2}\right)\,,
\eea
where the evaluation of the sub-leading term is the main result of this work.

\section{Discussion}

In this work we have presented a detailed evaluation of the next to leading order term in the mutual information of disks in a $(2+1)$ space-time dimensional free field theory.
This coefficient has not been evaluated before, neither analytically nor numerically, so there are no other results which we could compare it with. However following the spirit of  \cite{Agon:2015ftl} one could wonder whether this coefficient is in anyway related to the subleading coefficient  computed for CFT$_2$ in \cite{Perlmutter:2013paa} (with a $\Delta$ associated to the free field theory case for $3$ spacetime dimensions, that is $\Delta=1/2$). In fact, the numeric value of these numbers are surprisingly similar (for CFT$_2$ this is $8/15 \approx 0.53\hat{3}$, while for (2+1) free field theory is $6/15+4/(3\pi^2)\approx 0.5350\cdots$). Although, we do not expect any kind of universality for the next to leading term \footnote{As it occurs for the leading coefficient in this expansion. This is explained by the universality of the $2$ point functions in CFT's \cite{Agon:2015ftl}.}, it would be interesting to explore whether this approximate equality comes from some unknown bound that explains it.

We can estimate a hypothetical critical radius which might be interpreted as a break down of the power series expansion, by estimating the distance $r_c$ at which the leading and next to leading order terms are of the same order of magnitude. That is
\bea
\left(\frac{r_c^2}{R_AR_B}\right)\approx \left(\frac{6}{5}+\frac{4}{\pi^2}\right) \,.
\eea
Assuming $R_A=R_B=R$,  we get:
\bea
r_c\approx R\left(\frac{6}{5}+\frac{4}{\pi^2}\right)^{1/2}=1.267R \,.
\eea
This is clearly  beyond the validity of the expansion, since it conflicts with the geometric set up.  We interpret this as a justification of the expansion  (\ref{subs}) for large  $r\gg R$.

One of the motivations of this study was to explore the possibility of non-analyticity in the R\'enyi parameter n for the sub-leading term in the continuation  $n\rightarrow 1.$ There is no indication of non-analyticity in this continuation to the order in the large distance expansion we considered.

One can also consider the holographic expansion of the mutual information of two regions $A$ and $B$ on the boundary surface. The leading term at ${\cal  O}(N^2)$ in the $1/N$ or $G_N$ expansion vanishes, by the Ryu-Takayanagi argument \cite{Ryu:2006bv}. In this context the leading large distance term is ${\cal  O}(1)\sim{\cal  O}(G_N^0)$ \cite{Headrick:2010zt} \cite{Barrella:2013wja} \cite{Faulkner:2013ana}, and a phase-transition between bulk surfaces of different topologies occurs \cite{Headrick:2010zt}. These results present a challenge for the holographic calculation of mutual information to understand comparable terms in a large $r$ expansion\footnote{Notice that an exact match between the large distance holographic mutual information and its CFT dual was found by explicit calculation in \cite{Agon:2015ftl}, so it would be interesting to explore the next to leading terms in both the CFT and its bulk dual.}.

\section*{Acknowledgements}
We thank Matthew Headrick for several useful comments. H.J.S. and C. A. are supported in part by the DOE by grant DE-SC0009987. C. A is also supported in part by the National Science Foundation via CAREER Grant No. PHY10-53842 awarded to Matthew Headrick.
\appendix

\section{Calculation of single sum term \label{A1}}
We need to evaluate the expression (\ref{4.6}), this is
\bea
 &&\sum_{j=1}^{n-1}\langle\phi_j(1)\phi_0(1)\rangle^4 \nonumber \\
 &&=\frac{1}{16\pi^4}\int_0^\infty... \int_0^\infty\frac{(1-(xyzw)^{n-1})(1-x)(1-y)(1-z)(1-w)dxdydzdw }{(1+x)(1+y)(1+z)(1+w)(1-xyzw)(1-x^n)(1-y^n)(1-z^n)(1-w^n)}\,. \nonumber \\
 \eea
 In the $n\to 1$ limit, this is:
 \bea
 \sim -\frac{(n-1)}{16\pi^4}\int_0^\infty... \int_0^\infty\frac{\log(xyzw)dxdydzdw }{(1+x)(1+y)(1+z)(1+w)(1-xyzw)}\,.\nonumber \\
 \eea
 This integral can be done by first making the  consecutive and systematic change of variables:
 {$\{x'=xyzw, y=y, z=z, w=w\}$}, {$\{x'=x,y'=yzw, z=z,w=w\}$} and { $\{x=x, y'=y,z=zw, w=w\}$}
 \bea
 \sim -\frac{(n-1)}{16\pi^4}\int_0^\infty... \int_0^\infty\frac{\log(x)dxdydzdw }{(y+x)(z+y)(w+z)(1+w)(1-x)}\,.\nonumber \\
 \eea
 Doing the $w$ integral first we find
\bea
&\sim & \frac{(n-1)}{16\pi^4}\int_0^{\infty}dy\int_0^{\infty}\frac{dx\log x}{(1-x)(x+y)}\int_0^{\infty}\frac{dz \log z}{(1-z)(z+y)}\nonumber \\
&=&\frac{(n-1)}{16\pi^4}\int_0^{\infty}dy\left(-\frac{\pi^2+(\log y)^2}{2(1+y)}\right)^2=\frac{(n-1)}{30}\,.
\eea

\section{Calculation of the double sum term \label{A2}}
We are interested in evaluating the term from (\ref{4.12}), which is
\bea
&&\sum_{j=2}^{n-1}\langle\phi_{j}(1)\phi_0(1)\rangle^2\sum_{j'=1}^{j-1}\langle\phi_{j'}(1)\phi_0(1)\rangle^2\nonumber \\
&&=\frac{1}{16\pi^4}\sum_{j=2}^{n-1}\int_0^{\infty}\frac{dxdydzdw (1-(zw)^{j-1})(xy)^{j-1}(1-x)(1-y)(1-z)(1-w)}{(1-zw)(1+x)(1+y)(1+z)(1+w)(1-x^n)(1-y^n)(1-z^n)(1-w^n)}\,.\nonumber \\
\eea
One can add to this expression the term $j=1$ of the sum, since that will not contribute. Therefore carrying out the second sum and taking the $n\to 1$ limit we get:
\bea
=-\frac{(n-1)}{16\pi^4}\Bigg[\int_0^\infty... \int_0^\infty\frac{dxdydzdw \log(xy)}{(1-xy)(1-zw)(1+x)(1+y)(1+z)(1+w)}\nonumber \\
-\int_0^\infty... \int_0^\infty\frac{dxdydzdw \log(xyzw)}{(1-xyzw)(1-zw)(1+x)(1+y)(1+z)(1+w)}\Bigg]\,.
\eea
 We observe that these integrals diverge individually for $zw\to 1$, but not when combined. In order to cancel the spurious pole in $zw=1$ it is convenient to use the same change of variables used in the single sum term\footnote{In the first integral $\{x'=xy,y=y z'=zw, w=w\}$, while in the second $\{x'=xyzw, y'=yzw,z'=zw,w=w\}$}:
 \bea
=-\frac{(n-1)}{16\pi^4}\Bigg[\int_0^\infty... \int_0^\infty\frac{dxdydzdw \log(x)}{(1-x)(1-z)(y+x)(1+y)(w+z)(1+w)}\nonumber \\
-\int_0^\infty... \int_0^\infty\frac{dxdydzdw \log(x)}{(1-x)(1-z)(y+x)(z+y)(w+z)(1+w)}\Bigg]\,.
\eea
When added together, the term: $\frac{1}{1+y}-\frac{1}{z+y}=\frac{z-1}{(1+y)(z+y)}$ cancels the spurious pole.
\bea
=\frac{(n-1)}{16\pi^4}\Bigg[\int_0^\infty... \int_0^\infty\frac{dxdydzdw \log(x)}{(1-x)(y+x)(1+y)(z+y)(w+z)(1+w)}\Bigg]\,.
\eea
Again doing  the $w$ integral first, we get:
\bea
&&=-\frac{(n-1)}{16\pi^4}\Bigg[\int_0^\infty\frac{dy}{1+y} \int_0^\infty\frac{dz\log z}{(1-z)(y+z)}\int_0^\infty\frac{dx \log x}{(1-x)(y+x)}\Bigg]\nonumber \\
&&=-\frac{(n-1)}{16\pi^4}\int_0^\infty\frac{dy}{1+y}\left(-\frac{\pi^2+(\log y)^2}{2(1+y)}\right)^2=-\frac{(n-1)}{60}\,.
\eea

\section{Calculation of the triple sum term \label{A3}}

The triple sum term given by (\ref{4.18}) is
\bea
&& 4n\left(\frac{R_AR_B}{r^2}\right)^2\sum_{j=3}^{n-1}\sum_{k=2}^{j-1}
\sum_{l=1}^{k-1}\Big[\langle \phi_j(1)\phi_k(1)\rangle^2\langle \phi_l(1)\phi_0(1)\rangle^2+\langle \phi_k(1)\phi_l(1)\rangle^2\langle \phi_j(1)\phi_0(1)\rangle^2\nonumber \\
&& \qquad\qquad \qquad +\langle \phi_j(1)\phi_l(1)\rangle^2\langle \phi_k(1)\phi_0(1)\rangle^2\nonumber \\
&&\qquad \qquad\quad \quad +2\langle \phi_0(1)\phi_j(1)\rangle \langle \phi_j(1)\phi_l(1) \rangle\langle \phi_k(1)\phi_{l}(1)\rangle \langle \phi_k(1)\phi_0(1) \rangle\nonumber \\ 
&& \qquad\qquad \qquad \quad+2\langle \phi_0(1)\phi_k(1)\rangle \langle \phi_k(1)\phi_j(1) \rangle\langle \phi_j(1)\phi_{l}(1)\rangle \langle \phi_l(1)\phi_0(1) \rangle\nonumber \\
&&\qquad \qquad\qquad \qquad +2\langle \phi_0(1)\phi_l(1)\rangle \langle \phi_k(1)\phi_l(1) \rangle\langle \phi_j(1)\phi_{k}(1)\rangle \langle \phi_j(1)\phi_0(1) \rangle
\Big]\,. \nonumber \\
\eea
Writing all propagators in terms of integrals leads to
\bea
&& 4n\left(\frac{R_AR_B}{r^2}\right)^2\frac{1}{16\pi^4}\int_0^\infty... \int_0^\infty\frac{(1-x)(1-y)(1-z)(1-w)dxdydzdw }{(1+x)(1+y)(1+z)(1+w)(1-x^n)(1-y^n)(1-z^n)(1-w^n)} \nonumber \\ && \sum_{j=3}^{n-1}\sum_{k=2}^{j-1}
\sum_{l=1}^{k-1}\Big[(xy)^{l-1}(zw)^{j-k-1}+(zw)^{k-l-1}(xy)^{j-1}+(zw)^{j-l-1}(xy)^{k-1}\nonumber \\&&
+2x^{j-1}y^{j-l-1}z^{k-l-1}w^{k-1}+2x^{k-1}y^{j-k-1}z^{j-l-1}w^{l-1}+2x^{l-1}y^{k-l-1}z^{j-k-1}w^{j-1}
\Big]\,. \nonumber \\
\eea
The procedure to carry out all the sums consists in using the geometric formulas for the partial sums, complete the terms to get a sum which starts at $k=1$ and $j=1$ respectively, and again apply  the geometric formulas for the sums. The result is
\bea
\label{3sums}
&&4n\left(\frac{R_AR_B}{r^2}\right)^2\frac{1}{16\pi^4}\int_0^\infty... \int_0^\infty\frac{(1-x)(1-y)(1-z)(1-w)dxdydzdw }{(1+x)(1+y)(1+z)(1+w)(1-x^n)(1-y^n)(1-z^n)(1-w^n)} \nonumber \\ && \sum_{j=1}^{n-1}\Big\{\frac{1}{1-xy}\left[\frac{1-(zw)^{j-1}}{1-zw}-\frac{(zw)^{j-1}-(xy)^{j-1}}{zw-xy}\right] \nonumber \\
&&-\frac{1}{1-zw}\left[\frac{(zw)^{j-1}-(xyzw)^{j-1}}{1-xy}-zw \frac{(zw)^{j-1}-(xy)^{j-1}}{zw-xy}\right] \nonumber \\
&&-\frac{1}{1-zw}\left[\frac{(xy)^{j-1}-(xyzw)^{j-1}}{1-xy}-(j-1)(xy)^{j-1}\right] \nonumber \\
 &&+\frac{2}{1-yz}\left[y\frac{(xy)^{j-1}-(xw)^{j-1}}{y-w}- \frac{(xy)^{j-1}-(xyzw)^{j-1}}{1-zw}\right] \nonumber \\
 && +\frac{2z}{z-w}\left[\frac{(yz)^{j-1}-(xw)^{j-1}}{yz-xw}- \frac{(yz)^{j-1}-(xz)^{j-1}}{yz-xz}\right] \nonumber \\
&& +\frac{2}{x-y}\left[\frac{(wz)^{j-1}-(wx)^{j-1}}{z-x}- \frac{(wz)^{j-1}-(wy)^{j-1}}{z-y}\right]
\Big\} \,. \nonumber \\
\eea
After  careful inspection, it becomes evident that by combining the previous terms appropriately, the full integral does not have poles. We can  proceed systematically to make this fact explicit in order to carry out all the integrals. We found it convenient in that sense to combine the first three lines of (\ref{3sums}) and rewrite it as
\bea
&&4n\left(\frac{R_AR_B}{r^2}\right)^2\frac{1}{16\pi^4}\int_0^\infty... \int_0^\infty\frac{(1-x)(1-y)(1-z)(1-w)dxdydzdw }{(1+x)(1+y)(1+z)(1+w)(1-x^n)(1-y^n)(1-z^n)(1-w^n)} \nonumber \\ && \sum_{j=1}^{n-1}\Big\{\left(\frac{1-(zw)^{j-1}}{1-zw}\right)\left(\frac{1-(xy)^{j-1}}{1-xy}\right) \nonumber \\
&&-\frac{(xy)^{j-1}}{1-zw}\left[\left(\frac{1-(zw)^{j-1}}{1-zw}\right)-(j-1)\right]-\left(\frac{(zw)^{j-1}-(xy)^{j-1}}{zw-xy}\right)\nonumber \\
 &&+\frac{2}{1-yz}\left[y\frac{(xy)^{j-1}-(xw)^{j-1}}{y-w}- \frac{(xy)^{j-1}-(xyzw)^{j-1}}{1-zw}\right] \nonumber \\
 && +\frac{2z}{z-w}\left[\frac{(yz)^{j-1}-(xw)^{j-1}}{yz-xw}- \frac{(yz)^{j-1}-(xz)^{j-1}}{yz-xz}\right] \nonumber \\
&& +\frac{2}{x-y}\left[\frac{(wz)^{j-1}-(wx)^{j-1}}{z-x}- \frac{(wz)^{j-1}-(wy)^{j-1}}{z-y}\right]
\Big\} \,. \nonumber \\
\eea
Now  take the $n\to 1$ limit of the full expression:
\bea
\label{3sums1}
&&-4n(n-1)\left(\frac{R_AR_B}{r^2}\right)^2\frac{1}{16\pi^4}\int_0^\infty... \int_0^\infty\frac{dxdydzdw }{(1+x)(1+y)(1+z)(1+w)} \nonumber \\
&& \Big\{-\frac{1}{(1-zw)(1-xy)} \left(1+\frac{\log(xy)}{1-xy}+\frac{\log(zw)}{1-zw}-\frac{\log(xyzw)}{1-xyzw} \right) \nonumber \\
&&-\frac{1}{(1-zw)^2}\left(\frac{\log(xy)}{1-xy}-\frac{\log(xyzw)}{1-xyzw}\right)+\frac{xy}{1-zw}\frac{d}{d(xy)}\left(\frac{\log(xy)}{1-xy}\right) \nonumber \\
&&-\frac{1}{zw-xy}\left(\frac{\log(zw)}{1-zw}-\frac{\log(xy)}{1-xy}\right)\nonumber \\
 &&+\frac{2}{1-yz}\left[\frac{y}{y-w}\left(\frac{\log(xy)}{1-xy}-\frac{\log(xw)}{1-xw}\right)-\frac1{1-zw}\left(\frac{\log(xy)}{1-xy}-\frac{\log(xyzw)}{1-xyzw}\right)\right] \nonumber \\
 &&+\frac{2z}{w-z}\left[\frac{1}{yz-wx}\left(\frac{\log(yz)}{1-yz}-\frac{\log(xw)}{1-xw}\right)-\frac1{yz-xz}\left(\frac{\log(yz)}{1-yz}-\frac{\log(xz)}{1-xz}\right)\right] \nonumber \\
 && +\frac{2}{x-y}\left[\frac{1}{z-x}\left(\frac{\log(wz)}{1-wz}-\frac{\log(xw)}{1-xw}\right)-\frac1{z-y}\left(\frac{\log(wz)}{1-wz}-\frac{\log(wy)}{1-wy}\right)\right]
\Big\}\,. \nonumber \\
\eea
We now explain how to compute all the above integrals.

\subsection{Integral evaluation}
To carry out the multi-variable integrals of (\ref{3sums1}) it is convenient to analyze the integrals in appropriate pole-free combinations:
Let's consider the first three lines of integrals together:
\bea
&&\frac{1}{16\pi^4}\int_0^\infty... \int_0^\infty\frac{dxdydzdw }{(1+x)(1+y)(1+z)(1+w)} \nonumber \\
&& \Big\{-\frac{1}{(1-zw)(1-xy)} \left(1+\frac{\log(xy)}{1-xy}+\frac{\log(zw)}{1-zw}-\frac{\log(xyzw)}{1-xyzw} \right) \nonumber \\
&&-\frac{1}{(1-zw)^2}\left(\frac{\log(xy)}{1-xy}-\frac{\log(xyzw)}{1-xyzw}\right)+\frac{xy}{1-zw}\frac{d}{d(xy)}\left(\frac{\log(xy)}{1-xy}\right)\nonumber \\
&&
-\frac{1}{zw-xy}\left(\frac{\log(zw)}{1-zw}-\frac{\log(xy)}{1-xy}\right)\Big\}\,.
\eea
It is clear from the structure of the integrand that we can performed the next change of variables: $(x'=xy, y'=y, z'=zy,w'=w)$ followed by the relabel $(x=x',y=y',z=z',w=w')$ of all the integrals simultaneously. That leads to.
\bea
&&\frac{1}{16\pi^4}\int_0^\infty \int_0^\infty dx dz \int_0^\infty \frac{dy }{(y+x)(1+y)} \int_0^\infty \frac{dw }{(w+z)(1+w)} \nonumber \\
&& \Big\{-\frac{1}{(1-z)(1-x)} \left(1+\frac{\log x}{1-x}+\frac{\log z}{1-z}-\frac{\log(xz)}{1-xz} \right) \nonumber \\
&&-\frac{1}{(1-z)^2}\left(\frac{\log z}{1-z}-\frac{\log(xz)}{1-xz}\right)+\frac{x}{1-z}\frac{d}{dx}\left(\frac{\log x}{1-x}\right)\nonumber \\
&&
-\frac{1}{z-x}\left(\frac{\log z}{1-z}-\frac{\log(x)}{1-x}\right)\Big\}\,.
\eea
We  evaluate the $y$ and $w$ integrals, and end up with a finite and well defined double integral in $x$ and $z$
\bea
&&\frac{1}{16\pi^4}\int_0^\infty dx \left(\frac{\log x}{1-x}\right)\int_0^\infty dz \left(\frac{\log z}{1-z}\right) \Big\{-\frac{1}{z-x}\left(\frac{\log z}{1-z}-\frac{\log(x)}{1-x}\right)\nonumber \\
&&-\frac{1}{(1-z)(1-x)} \left(1+\frac{\log x}{1-x}+\frac{\log z}{1-z}-\frac{\log(xz)}{1-xz} \right) \nonumber \\
&&-\frac{1}{(1-z)^2}\left(\frac{\log z}{1-z}-\frac{\log(xz)}{1-xz}\right)+\frac{x}{1-z}\frac{d}{dx}\left(\frac{\log x}{1-x}\right) \Big\}=-\frac{1}{30}\,.
\eea
This integral can be reduced to a single integral in terms of polynomials of \{$x$, log functions as well as poly-logarithmic functions \}. The exact value was derived from a numerical evaluation.
The remaining three lines of integrals in (\ref{3sums1}) can be evaluated in a more systematic way by removing  the spurious poles one by one,  in the integrals
\bea
\label{Bterms}
&&\frac{1}{16\pi^4}\int_0^\infty... \int_0^\infty\frac{dxdydzdw }{(1+x)(1+y)(1+z)(1+w)} \nonumber \\
&&+\frac{2}{1-yz}\left[\frac{y}{y-w}\left(\frac{\log(xy)}{1-xy}-\frac{\log(xw)}{1-xw}\right)-\frac1{1-zw}\left(\frac{\log(xy)}{1-xy}-\frac{\log(xyzw)}{1-xyzw}\right)\right] \nonumber \\
 &&+\frac{2z}{w-z}\left[\frac{1}{yz-wx}\left(\frac{\log(yz)}{1-yz}-\frac{\log(xw)}{1-xw}\right)-\frac1{yz-xz}\left(\frac{\log(yz)}{1-yz}-\frac{\log(xz)}{1-xz}\right)\right] \nonumber \\
 && +\frac{2}{x-y}\left[\frac{1}{z-x}\left(\frac{\log(wz)}{1-wz}-\frac{\log(xw)}{1-xw}\right)-\frac1{z-y}\left(\frac{\log(wz)}{1-wz}-\frac{\log(wy)}{1-wy}\right)\right]
\Big\}\,. \nonumber \\
\eea
We  consider this line by line, and do a different change of variables in each individual integral inside a given line. For example, in
\bea
&&\frac{1}{16\pi^4}\int_0^\infty... \int_0^\infty\frac{dxdydzdw }{(1+x)(1+y)(1+z)(1+w)} \nonumber \\
&&\frac{2}{1-yz}\left[\frac{y}{y-w}\left(\frac{\log(xy)}{1-xy}-\frac{\log(xw)}{1-xw}\right)-\frac1{1-zw}\left(\frac{\log(xy)}{1-xy}-\frac{\log(xyzw)}{1-xyzw}\right)\right]\Big\}\,.\nonumber \\
\eea
We set $(x'=xy, y'=y, ...)$ in the first and third integrals,  $(x'=xw, w'=w, ...)$ in the second, and $(x'=xyzw, y'=y, ...)$ in the fourth one. After removing the primes  we have
\bea
&&\frac{1}{16\pi^4}\int_0^\infty... \int_0^\infty\frac{dydzdw }{(1+y)(1+z)(1+w)}\int_0^{\infty}dx\left(\frac{\log x}{1-x}\right) \nonumber \\
&&\frac{2}{1-yz}\left[\frac{y}{y-w}\left(\frac{1}{y+x}-\frac{1}{w+x}\right)-\frac1{1-zw}\left(\frac{1}{y+x}-\frac{1}{yzw+x}\right)\right]\,,\nonumber \\
\eea
with these simple steps we cancel all the poles explicitly by simply adding the different fractions
\bea
&&\frac{1}{8\pi^4}\int_0^\infty... \int_0^\infty\frac{dydzdw }{(1+y)(1+z)(1+w)}\int_0^{\infty}dx\left(\frac{\log x}{1-x}\right)\frac{yw}{(y+x)(w+x)(yzw+x)}\,.\nonumber \\
\eea
As a further simplification we do the next change of variables: $(z'=zyw, y'=yw, x'=x, w'=w)$, and carry out the integrals on $w$ and $z$ respectively
\bea
&&\frac{1}{8\pi^4}\int_0^{\infty}dx\left(\frac{\log x}{1-x}\right)\int_0^{\infty}dy\int_0^\infty\frac{dw \, \, yw}{(1+w)(w+y)(y+xw)(w+x)}\int_0^\infty\frac{dz }{(y+z)(z+x)}\nonumber \\
&&=\frac{1}{8\pi^4}\int_0^{\infty}\int_0^{\infty}\frac{dx dy\, y}{(x-y)(x^2-y)}\left(\frac{\log x}{1-x}\right)\left(\frac{\log x-\log y}{x-y}\right)\left(x\frac{\log (x^2)}{1-x}-(x+y)\frac{\log y}{1-y}\right)\nonumber \\
&&=-\frac{1}{45}\left(\frac{1}{2}+\frac{15}{4\pi^2}\right)=-\left(\frac{1}{90}+\frac{1}{12\pi^2}\right)\,.
\eea
After following the same steps as the previous evaluation, we get for the final integral expressions for the second and third lines of (\ref{Bterms}):
\bea
&&-\frac{1}{8\pi^4}\int_0^{\infty}\int_0^{\infty}\frac{dx dy}{(x-y)^2}\left(\frac{\log x}{1-x}\right)\left(\frac{\log x}{1-x}-\frac{\log y}{1-y}\right)\left(x\frac{\log x}{1-x}-y\frac{\log y}{1-y}\right)\nonumber \\
&&=-\frac{1}{45}\left(\frac{1}{2}+\frac{15}{4\pi^2}\right)=-\left(\frac{1}{90}+\frac{1}{12\pi^2}\right)
\eea
 and
\bea
-\frac{1}{8\pi^4}\int_0^{\infty}dw\left(\frac{\log w}{1-w}
\right)^4=-\frac{1}{45}\left(1+\frac{15}{2\pi^2}\right)=-\left(\frac{1}{45}+\frac{1}{6\pi^2}\right)
\eea
 respectively. The final value of  the integral  in (\ref{3sums1}) is:
 \bea
 -\left(\frac{1}{30}+\frac{2}{45}+\frac{1}{3\pi^2}\right)\,.
 \eea
With this result at hand we can write the final result for the triple sum term, which is
\bea
\frac{n}{ 4!r^4}\sum_{j\neq j'\neq l}^{n-1}C^A_{jj'l0}C^B_{jj'l0}=
{4n(n-1)}\left(\frac{R_AR_B}{r^2}\right)^2\left(\frac{1}{30}+\frac{2}{45}+\frac{1}{3\pi^2}\right)\,.
\eea

\newpage
\bibliographystyle{utphys}

\bibliography{emirefs}

\end{document}